# Development of thin hydrogenated amorphous silicon detectors on a flexible substrate

M. Menichelli[1,*], M. Bizzarri[1,2], L. Calcagnile[3], M. Caprai[1], A.P. Caricato[3], R. Catalano[4], G.A.P. Cirrone[4], T. Croci[1,5], G. Cuttone[4], S. Dunand[6], M. Fabi[7,8], L. Frontini[9], B. Gianfelici[1,2], C. Grimani[7,8], M. Ionica[1], K. Kanxheri[1], M. Large[11], V. Liberali[9], M. Martino[3], G. Maruccio[3], G. Mazza[12], A. G. Monteduro[3], A. Morozzi[1], F. Moscatelli[1,13], S. Pallotta[7,14], A. Papi[1], D. Passeri[1,5] *Senior Member IEEE*, M. Pedio[1,13], M. Petasecca[11] *Member, IEEE*, G. Petringa[4], F. Peverini[1,2], L. Piccolo[12], P. Placidi[1,5], G. Quarta[3], S. Rizzato[3], G. Rossi[1], F. Sabbatini[7,8], A. Stabile[9], L. Servoli[1], C. Talamonti[7,14], M. Villani[7,8], R.J. Wheadon[12], N. Wyrsch[6].

*Abstract*— The HASPIDE (Hydrogenated Amorphous Silicon PIxels DEtectors) project aims at the development of thin hydrogenated amorphous silicon (a-Si:H) detectors on flexible substrates (mostly Polyimide) for beam monitoring, neutron detection and space applications. Since a-Si:H is a material with superior radiation hardness, the benefit for the above-mentioned applications can be appreciated mostly in radiation harsh environments. Furthermore, the possibility to deposit this material on flexible substrates like Polyimide (PI), polyethylene naphthalate (PEN) or polyethylene terephthalate (PET) facilitates the usage of these detectors in medical dosimetry, beam flux and beam profile measurements. Particularly interesting is its use when positioned directly on the flange of the vacuum-to-air separation interface in a beam line, as well as other applications where a thin self-standing radiation flux detector is envisaged. In this paper, the HASPIDE project will be described and some preliminary results on PI and glass substrates will be reported.

This work was partially supported by the "Fondazione Cassa di Risparmio di Perugia" RISAI project n. 2019.0245.

[1.] INFN, Sez. di Perugia, via Pascoli s.n.c. 06123 Perugia (ITALY)
[2.] Dip. di Fisica e Geologia dell'Università degli Studi di Perugia, via Pascoli s.n.c. 06123 Perugia (ITALY)
[3.] INFN and Dipartimento di Fisica e Matematica dell'Università del salento, Via per Arnesano, 73100 Lecce (ITALY)
[4.] INFN Laboratori Nazionali del Sud, Via S.Sofia 62, 95123 Catania (ITALY)
[5.] Dip. di Ingegneria dell'Università degli studi di Perugia, via G.Duranti 06125 Perugia (ITALY)
[6.] Ecole Polytechnique Fédérale de Lausanne (EPFL), Institute of Electrical and Microengineering (IME), Rue de la Maladière 71b, 2000 Neuchâtel, (SWITZERLAND).
[7.] INFN Sez. di Firenze, Via Sansone 1, 50019 Sesto Fiorentino (FI) (ITALY)
[8.] DiSPeA, Università di Urbino Carlo Bo, 61029 Urbino (PU) (ITALY)
[9.] INFN Sezione di Milano Via Celoria 16, 20133 Milano (ITALY)
[10.] Azienda ospedaliera S.Maria, via Tristano di Joannuccio, 05100 Terni (ITALY)
[11.] Centre for Medical Radiation Physics, University of Wollongong, Northfields Ave Wollongong NSW 2522, (AUSTRALIA).
[12.] INFN Sez. di Torino Via Pietro Giuria, 1 10125 Torino (ITALY)
[13.] CNR-IOM, via Pascoli s.n.c. 06123 Perugia (ITALY)
[14.] Dipartimento di Fisica Scienze Biomediche sperimentali e Cliniche "Mario Serio", Viale Morgagni 50, 50135 Firenze (FI) (ITALY)
* Corresponding author, Mauro.menichelli@pg.infn.it

*Index Terms*— Hydrogenated Silicon detectors, Radiation Hardness, Flexible detectors.

## I. INTRODUCTION

THERE is an increasing demand for radiation-resistant detectors capable of high dynamic range and precise measurement of fluxes of ionizing radiation of different kinds (photons, electrons, protons, ions). Current and new clinical procedures need high fluxes of particles, e.g. IORT [1] and FLASH therapy [2-4]. The development of new accelerator techniques, such as Reaccelerated Ion Beams (RIBs) [5] and laser driven accelerators [6] motivate the demand for specific detectors. In addition, these types of detectors can provide benefits for beam monitoring in nuclear physics, astrophysics, and medical research accelerators. Hence, the development of new detectors for dosimetry and ionizing radiation flux measurement is desirable.

One of the main challenges is to find detector materials that are useful for the production of large dynamic range detectors for dose-rates spanning several orders of magnitude, thin and therefore highly transparent to the beam, radiation-resistant, mechanically flexible and working both in air and vacuum.

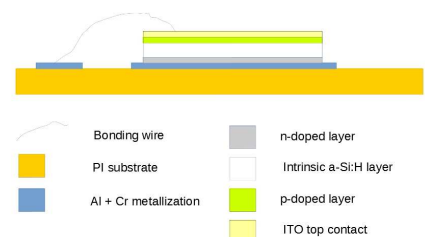

Fig.1 Structure of a p-i-n device deposited on PI

Hydrogenated Amorphous Silicon (a-Si:H) is a candidate material due to its intrinsic non-crystalline nature and alloying with H for effective defect passivation that offers a natural resistance to radiation damage and the possibility to grow it in a very thin layer and over large areas on a variety of substrates. The current technology is mature considering its widespread use in the solar cells and medical imaging industry. The supporting layer could be mechanically flexible like, for example, Polyimide (PI), allowing devices to be adaptable to a variety of shapes and surfaces. Plasma Enhanced Chemical Vapor Deposition (PECVD) of doped and intrinsic a-Si:H on metalized pads is a mature technology and a widely adopted method [7-9] used to fabricate, up to now, the majority of a-Si:H radiation detectors [10-11]. Other possible technologies are: Hot wire deposition (CAT-CVD), Pulsed Laser Deposition (PLD) [12] and reactive sputtering [13] which may complement PECVD in certain areas. Concerning the detector structure there are two possible options envisaged for this project. The first option is the more canonical p-i-n diode structure with n-doped a-Si:H deposited on the metal (Chromium + Aluminum) pad that is deposited on the substrate according to the scheme shown in fig.1. The second option is a charge selective contact diode device, the structure of a charge selective contacts device is similar to p-i-n device where instead of p-doped a-Si:H there is hole selective metal oxide (namely $MoO_x$) and instead of n-doped a-Si:H there is electron selective metal oxide (namely ZnO:Al or $TiO_2$). Charge selective contacts are based, not on the concentration of majority charge carriers densities (like in p-doped or n-doped a-Si:H) but on the different mobilities of the different charge carriers; if $\mu_e$ is the mobility of electrons and $\mu_h$ is the mobility of holes, in hole selective contacts $\mu_h \gg \mu_e$ while in electron selective contacts $\mu_e \gg \mu_h$. Radiation detectors fabricated with this technique have been recently tested for the first time in the framework of the 3D-SiAm experiment [14]. In Fig.2 a charge selective contact device manufactured with a 2 x 2 array of 4 x 4 $mm^2$ pixels is shown. In Fig. 3, a current versus voltage curve for two charge selective contact devices is displayed where it is possible to notice the rectifying behavior of this device design.

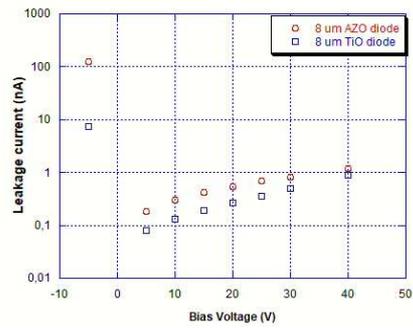

Fig. 3 Current versus Voltage characteristics of a 2 different charge selective contact devices one sample has $TiO_2$ electron selective contact and the other has ZnO:Al electron selective contact the rectifying behavior is evident in both devices

## II. PRELIMINARY TESTING OF P-I-N DEVICES ON PI

In the prototype detector shown in Fig. 4 the top p-type silicon layer is in coated with (indium Tin Oxide (IT) as top contact. These prototypes are then cut into 4 smaller arrays before being glued and bonded on a PCB frame (Fig.5). The larger diode (5 mm x 5 mm surface and 2.5 μm thickness a-Si:H deposited on PI) has been tested for leakage current and photo-current stimulated with a 30 kV x-ray beam at various currents (20-200 μA) from an X-ray tube (10W power, Vmax = 50kV, Imax = 200 μA) by Newton Scientific. Fig. 6 shows the results of an I/V scan from 2 to 12V bias performed at a controlled temperature (20 °C), where the maximum leakage current density was found to be 60 $nA/cm^2$ and at the selected operating Voltage of 8V is 28 $nA/cm^2$. The detector was then irradiated with the photocurrent response of the device shown in Fig. 7 at 0V and 8V detector bias. The linear regression coefficient was calculated as R = 0.996 with a current sensitivity of 0.80 ± 0.02 nC/cGy for 0V (Fig.7a) detector bias, and R = 0.994 and 6.3 ±0.8 nC/cGy for 8V detector bias (Fig.7b), respectively.

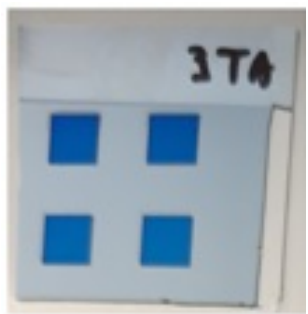

Fig.2 Prototype of charge selective contacts 2 x 2 device (each having 4 x 4 $mm^2$ area) array deposited on glass

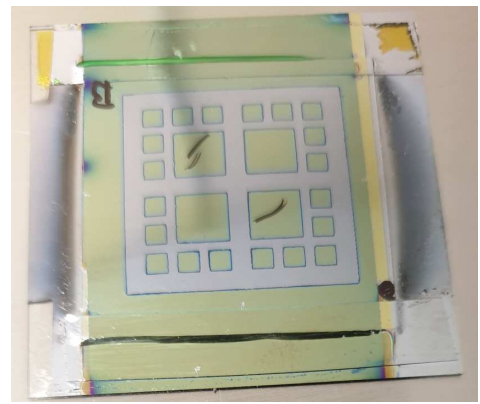

Fig.4 Picture of several devices deposited on PI the light green area is the ITO contacts the grey area is the intrinsic a-Si:H and the shiny grey area is the Cr+Al back contact.

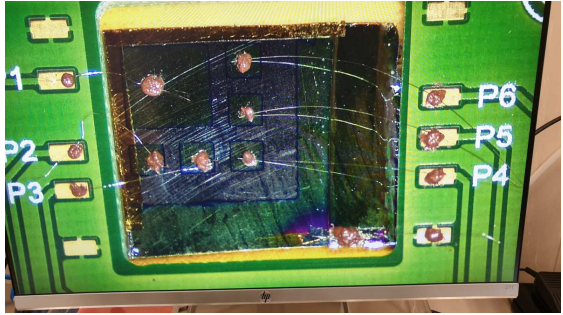

Fig.5 One quarter of the array shown in Fig.2 glued and bonded on a PCB frame

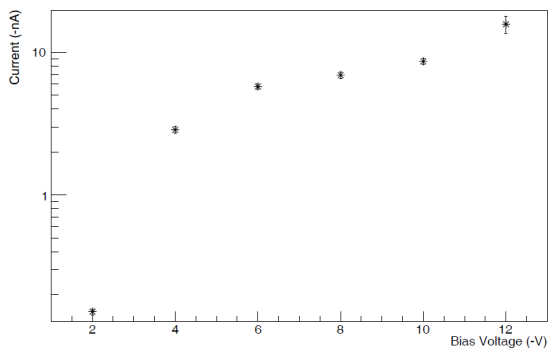

Fig.6 Leakage current in the bias voltage range 2-12V for a 5 mm x 5 mm a-Si:H p-i-n device on PI substrate.

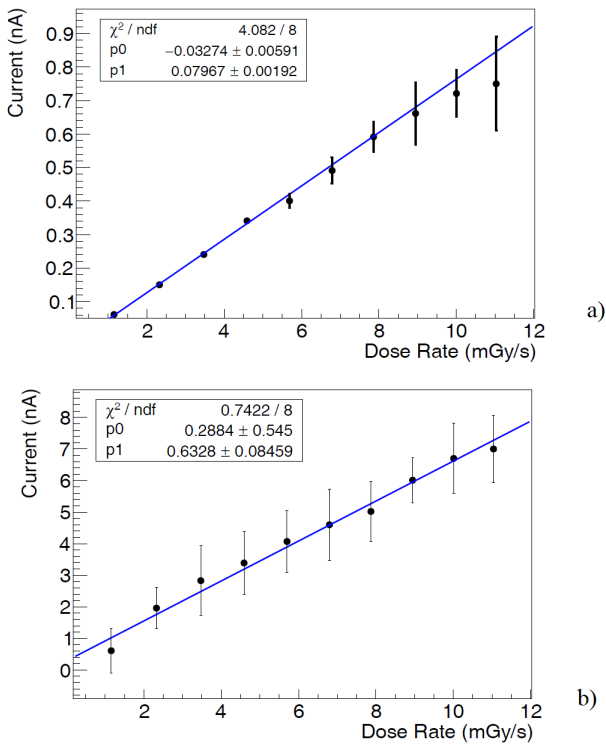

Fig. 7 Photo-current versus dose rate on a 5 mm x 5 mm pixel of the a-Si:H p-i-n device on PI, irradiated with an x-ray tube at 30 kV bias. The detector was biased at 0V in Fig.7a and 8V in Fig.7b.

## III. DETECTOR APPLICATIONS

As stated in the introduction, three applications were foreseen for these detectors:

- Beam monitoring for clinical (and also non-clinical) flux/dose measurements
- Space applications like medium-strong solar energetic particle event monitoring
- Neutron detection

Beam monitoring applications do not include only the measurement of the beam flux (including beam profile definition and dose rate measurements), but also real-time measurements of the beam incident on a patient's body. For these two aspects, the application of a radiation hard, flexible and radiation transparent detector will be very beneficial. Concerning transparency, an ideal TRansmission Detector (TRD) device should monitor the radiation fluence with high accuracy (2% or less) without perturbing the incoming radiation beam. The main detectors currently available are 2D detector arrays based on ionization chambers [15-16] or silicon diodes [17-19]. Approximately 1.1% of a 6 MV energy beam is absorbed by the diodes array, compared to the 5% for the ion-chamber-based TRDs. Furthermore, there is a lower rate of secondary electrons (smaller electron contamination) produced within the a-Si:H devices due to the lack of high Z materials. With these technologies, transmission factors must be calculated for each energy option and each treatment machine. The rigid shape of the detectors and their bulkiness constitute a problem too. If a flexible and thin sensor could be available for relative dosimetry, it would represent a huge technological advancement.

Space applications for a-Si:H detectors are mainly meant to solar flare and solar energetic particle (SEP) monitoring up to 400-600 MeV. Large Space Weather phenomena pose a health risk to astronauts and affect electronics and instrument performance on board space missions. SEP flux multi-spacecraft observations at different distances from the Sun are mandatory to study how the Sun generates and accelerates SEPs and how SEPs propagate in the interplanetary plasma [20-23]. Despite ESA and NASA fly particle detectors on a large fleet of spacecraft, the totality of these instruments (including Solar Orbiter and Parker Solar Probe presently orbiting near the Sun) limit the detection of solar particles below 100-200 MeV. Moreover, these missions are not optimized to issue space weather alerts [24-25] for strong geomagnetic storms [26] occurrence. The availability of small-size sensors capable of real-time measurements also on small near-Earth satellites (better in the region of the South Atlantic Anomaly) would boost the comprehension of SEPs and early warning capability.

Neutron detection plays an important role in different disciplines in basic and applied research, in industry and medicine, such as homeland security and nuclear safeguards, material investigation based on *n-scattering*, *n-monitoring* in nuclear power plants and fusion research, cancer therapy and cultural heritage [27-30]. Since neutrons do not have an electrical charge, they do not produce direct ionization while passing through matter. Their detection is then carried out by employing a double-step process from neutron to secondary charged particles and subsequent detection by ionization. In recent years considerable research efforts have been put into developing new conversion materials due to the shortage of $^3$He, which constituted the traditional solution to the problem. The two most explored alternatives are based on $^6$Li and $^{10}$B [31-37]. In particular, the deposition of $^{10}$B via Pulsed Laser Deposition is attractive for its fabrication on flexible substrates kept at room temperatures such as PMMA, Kapton and plastic scintillators. These depositions may be adopted on a 15 μm thick a-Si:H device detecting the alpha particle generated in the neutron conversion. Furthermore, the use of boron targets isotopically enriched in $^{10}$B (>96%) allowed the enhancement of the detection efficiency [36].

The main advantages of the PLD technique to deposit thin films are:
• possibility to change many independent parameters;
• the high energy of the ablated/ejected particles;
• deposition on thermolabile substrates;
• good adhesion on many kinds of substrates without the need for particular substrate treatment.

IV. CONCLUSIONS

The HASPIDE project aims at the construction of flexible planar detectors for radiation flux measurements and neutron detection. For these purposes, the experiment is also developing two different readout electronics: the first implies current measurement based on flux measurements and the second is a charge integrator based on single particle detection. The first samples based on p-i-n structure deposited via PECVD on PI are presently available and preliminary characterization and flux measurement results are presented in this paper. Although the leakage current of these early samples is quite high the response to x-ray fluxes in the range between 1.1 to 11 mGy/s is very linear with good sensitivities (0.8 nC/cGy at 0V bias and 6.3 nC/cGy at 8V). HASPIDE is a 3 year program funded by INFN that will explore the potential of these detectors in the field on beam flux monitoring and dosimetry, space applications and neutron detection.